\documentclass[aps,prmaterials,reprint,longbibliography,superscriptaddress,floatfix]{revtex4-2}%

\usepackage{amsmath,amssymb}
\usepackage{xspace} 
\usepackage{graphicx}

\usepackage{color} 
\usepackage{dcolumn}
\usepackage{bm}

\newcommand{\BaNiP}{BaNi$_2$P$_4$}

\begin{document}

\title{Superlinear Temperature-Dependent Resistivity and Structural Phase Transition in \BaNiP}

\author{E. H. Krenkel}
\email[Current affiliation: ]{Los Alamos National Laboratory}
\affiliation{Ames National Laboratory, Iowa State University, Ames, Iowa 50011, USA}
\affiliation{Department of Physics and Astronomy, Iowa State University, Ames, Iowa 50011, USA}

\author{M. A. Tanatar}
\email[Corresponding author: ]{tanatar@ameslab.gov}
\affiliation{Ames National Laboratory, Iowa State University, Ames, Iowa 50011, USA}
\affiliation{Department of Physics and Astronomy, Iowa State University, Ames, Iowa 50011, USA}

\author{E. I. Timmons}
\email[Current affiliation: ]{Emerson Innovation Center, Fisher Technology}
\affiliation{Ames National Laboratory, Iowa State University, Ames, Iowa 50011, USA}
\affiliation{Department of Physics and Astronomy, Iowa State University, Ames, Iowa 50011, USA}

\author{S. L. Bud’ko}
\affiliation{Ames National Laboratory, Iowa State University, Ames, Iowa 50011, USA}
\affiliation{Department of Physics and Astronomy, Iowa State University, Ames, Iowa 50011, USA}

\author{P. C. Canfield}
\affiliation{Ames National Laboratory, Iowa State University, Ames, Iowa 50011, USA}
\affiliation{Department of Physics and Astronomy, Iowa State University, Ames, Iowa 50011, USA}

\author{Qing-Ping Ding}
\affiliation{Ames National Laboratory, Iowa State University, Ames, Iowa 50011, USA}

\author{Yuji Furukawa}
\affiliation{Ames National Laboratory, Iowa State University, Ames, Iowa 50011, USA}
\affiliation{Department of Physics and Astronomy, Iowa State University, Ames, Iowa 50011, USA}

\author{Lin-Lin Wang}
\affiliation{Ames National Laboratory, Iowa State University, Ames, Iowa 50011, USA}

\author{M. Ko\'nczykowski}
\affiliation{Laboratoire des Solides Irradi\'es, CEA/DRF/IRAMIS,\'Ecole Polytechnique, CNRS, Institut Polytechnique de Paris, 91128 Palaiseau, France }

\author{R. Grasset}
\affiliation{Laboratoire des Solides Irradi\'es, CEA/DRF/IRAMIS,\'Ecole Polytechnique, CNRS, Institut Polytechnique de Paris, 91128 Palaiseau, France }

\author{J.~ L. Niedziela}
\affiliation{Nuclear Nonproliferation Division, Oak Ridge National Laboratory, Oak Ridge, TN 37831, USA}

\author{Olivier Delaire}
\affiliation{Department of Mechanical Engineering and Materials Science, Department of Physics and Department of Chemistry, Duke University, Durham, NC 27708, USA}

\author{Gayatri Viswanathan}
\affiliation{Ames National Laboratory, Iowa State University, Ames, Iowa 50011, USA}
\affiliation{Department of Chemistry, Iowa State University,
Ames, Iowa 50011, USA}

\author{Jian Wang}
\affiliation{Department of Chemistry and Biochemistry, Wichita State University,
Wichita, Kansas  67260, USA}

\author{Kirill Kovnir}
\affiliation{Ames National Laboratory, Iowa State University, Ames, Iowa 50011, USA}
\affiliation{Department of Chemistry, Iowa State University,
Ames, Iowa 50011, USA}

\author{Ruslan Prozorov}
\affiliation{Ames National Laboratory, Iowa State University, Ames, Iowa 50011, USA}
\affiliation{Department of Physics and Astronomy, Iowa State University, Ames, Iowa 50011, USA}

\date{30 March 2026}

\begin{abstract}

The mechanism of anomalous superlinear temperature-dependent resistivity, $\rho (T)$, in the metallic unconventional clathrate BaNi$_2$P$_4$ was studied by examining its evolution with artificial disorder induced by low-temperature ($\sim$ 20 K) 2.5 MeV electron irradiation. We find a dominant effect of the tetragonal-orthorhombic transition at $T_s$ ($ \sim$373 to 378 K, depending on heat cycle rate and direction) on $\rho (T)$, with standard metallic $T-$linear resistivity above the transition and anomalous behavior in the orthorhombic phase below. The transition is accompanied by the formation of structural domains and a notable (about 4~K) hysteresis in the magnetization and resistivity measurements, clearly showing its first order character.  Matthiessen rule is obeyed both above and below the transition, suggesting negligible changes in the electronic structure. This conclusion is supported by the smooth evolution of the Hall effect through the transition. The Hall number is in good agreement with band structure calculations both above and below the transition. The transition temperature is notably suppressed with electron irradiation. Raman scattering at temperatures above room temperature find softening of local Ba vibration mode in the orthorhombic phase on approaching the transition.  $^{31}P$ NMR line splits in the orthorhombic phase, suggesting a partial shift of the Ba atom from the central position in the cage. We suggest that local Ba rattling leads to enhanced residual contribution to resistivity in the high temperature tetragonal phase, the decay of which is responsible for the anomalous temperature-dependent resistivity in the orthorhombic phase.  

\end{abstract}


\maketitle

\section{Introduction}





Clathrates are composed of large, cage-like structures which are only weakly coupled to the atoms in the center of their cages. The central atoms are able to rattle incoherently from each other, stymieing the formation of long-wavelength phonons which carry much of the thermal energy of the material, see \cite{review1,review2} for a review.
Because of this, clathrate materials have recently attracted much attention as potentially good thermoelectrics \cite{Kirill}.   In addition to low thermal conductivity $\kappa$, they also possess high electrical conductivity $\sigma$, \cite{first} an important combination for a high thermoelectric figure of merit $ZT=S^2T\sigma/\kappa$, where $S$ is the Seebeck coefficient. Another attractive feature of clathrates is the tunability of their structure, with easily-changeable symmetry of the central atom.

\begin{figure}[b]
\includegraphics[width=1.0\linewidth]{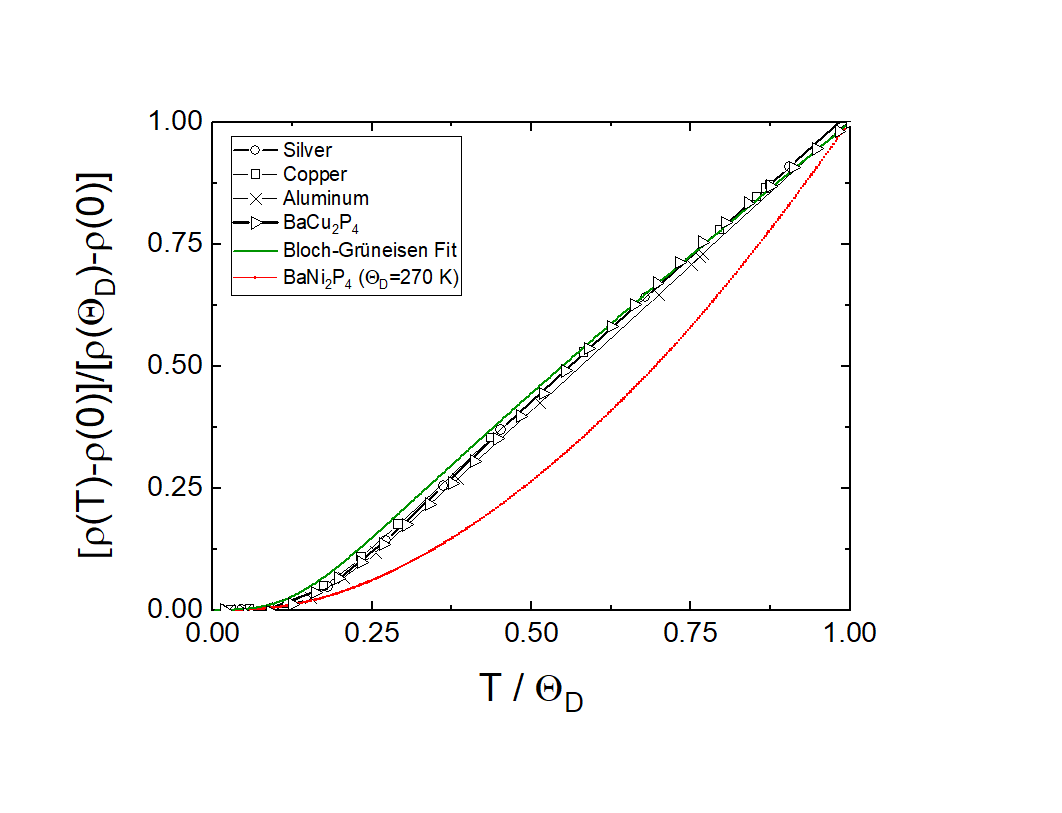}
\caption{Universal scaling of electrical resistivity of conventional metals with the Debye temperature \cite{Bardeen1940}. The plot presents electrical resistivity data for silver, copper and aluminum using normalized resistivity, $\rho/\rho(\Theta_D$), and temperature, $T/\Theta_D$, scales. The right triangles show resistivity data for clathrate BaCu$_2$P$_4$, following generic metallic behavior. Red curve shows the data for \BaNiP~ strongly deviating from the normal metallic temperature dependence. }
\label{fig:DebyeTemp}
\end{figure}

In the clathrate material  \BaNiP, while anomalous thermal behavior with small phonon contribution is expected, there is little in the band structure to suggest that it would not act as a good metal. The Fermi surface of \BaNiP~ suggests a good metallic phase, and the frequencies observed in de Haas-van Alphen oscillations are in reasonable agreement with the calculated band structure \cite{first}. Yet the temperature-dependent resistivity of this material is quite non-trivial, see Fig.~\ref{fig:DebyeTemp}. In good metals $\rho(T)$ follows Bloch-Gr\"uneisen (BG) theory in the form $\rho = \rho_0 + \rho_{BG}$, where $\rho_0$ is the scattering based on impurities and $\rho_{BG}$ is the scattering from electron scattering on phonons \cite{ZimanJ2001:}.  The form of $\rho_{BG}$ is given by 

\begin{equation}
\rho_{BG} \propto \frac{T^5}{\Theta_D} \int_0^{\theta_D / 2} \frac{x^5 dx}{(e^x - 1)(1-e^{-x})}
\label{Eq1BG}
\end{equation}

\noindent
where $\Theta_D$ is approximately equal to the Debye temperature from heat capacity measurements \cite{Bardeen1940}.  When this function is evaluated using the appropriate approximations, it becomes clear that its behavior is approximately linear above $1/5 \Theta_D$.  This is experimentally borne out in many metals, as shown in Fig.~\ref{fig:DebyeTemp}.  However, for BaNi$_2$P$_4$, this is notably violated.  As shown in Fig.~\ref{fig:DebyeTemp} and previously in \cite{first}, the resistivity of BaNi$_2$P$_4$ is superlinear up to very high temperatures. The Debye temperature of the compound determined from anisotropic displacement parameters in X-ray measurements of 237~K and from heat capacity measurements (205~K)  \cite{first}, suggests that $T-$linear variation should be observed starting from above 50~K or so. Interestingly, this temperature-dependent resistivity is clearly distinct from the normal, nearly linear  resistivity as determined on closely related clathrate materials BaCu$_2$P$_4$ single crystals (right triangles in Fig.~\ref{fig:DebyeTemp}) and SrNi$_2$P$_4$ \cite{SrNi2P4} and EuNi$_2$P$_4$ \cite{EuNi2P4} polycrystalline samples.

\begin{figure}[t]
    \centering
    \includegraphics[width = \linewidth]{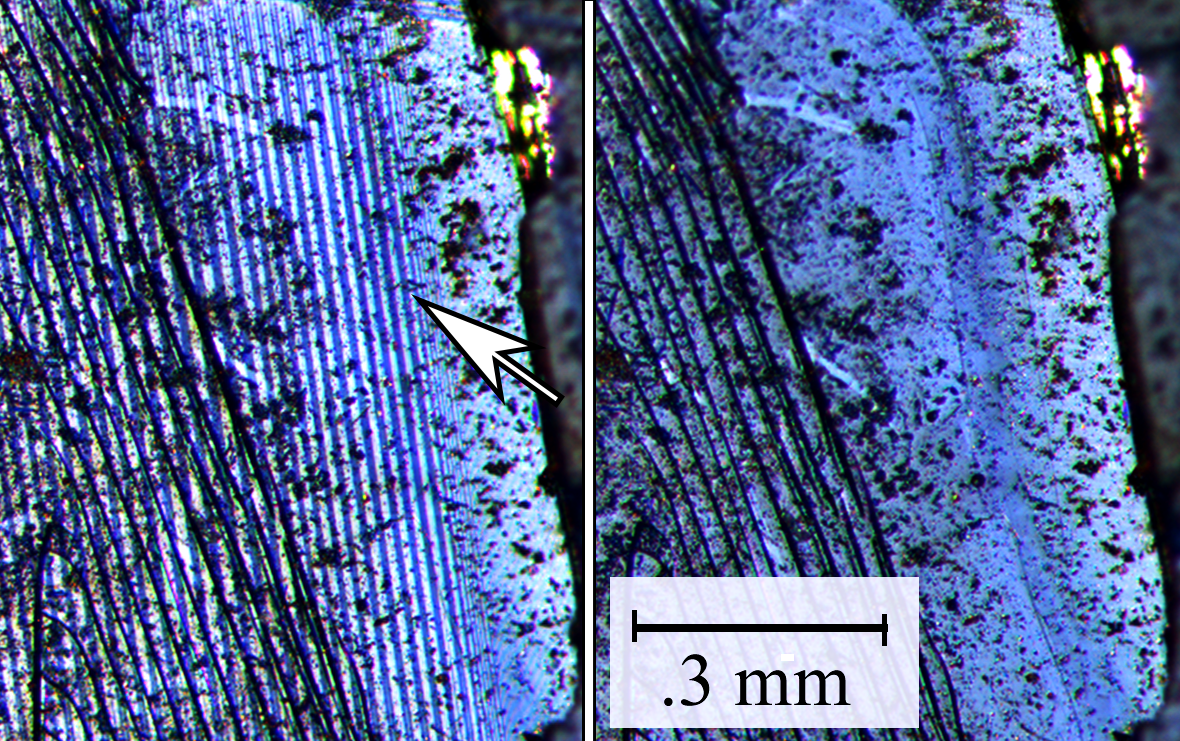}
    \caption{Polarized optics image of the orthorhombic domains in the crystal of \BaNiP~  at room temperature (left panel). Domains are seen as bright stripes crossing through the terraces on sample surface. On warming to the temperatures ($\sim$450~K) higher that structural transition temperature(376 ~K), the domains disappear leaving terraces unaffected  (right panel).}
    \label{fig:Polarized_Optics}
\end{figure}

It is known that \BaNiP undergoes a structural phase transition from a high temperature tetragonal (HTT) to a low temperature orthorhombic (LTO) phase at approximately 373 K \cite{Peierls}. The transition was initially assigned to the Peierls instability of the electronic system \cite{Peierls}. Later, it was argued that the transition is caused by the energy of the filled states, analogous to a collective Jahn-Teller distortion \cite{Alemany}. As a consequence of the orthorhombic distortion, the sample splits at room temperature into domains, as can be seen in the polarized optical microscope image in the left panel of Fig.~\ref{fig:Polarized_Optics}. On warming  above the transition temperature, the domains disappear.

Considering both the possibility of an anomalous phonon spectrum which can contribute to the anomalous temperature-dependent resistivity, and the possible effect of the structural transition on band structure, here we present study of the electronic transport properties of \BaNiP~ single crystals. We use electron irradiation to control the disorder scattering $\rho_0$.  We show that the transition does not lead to a notable change in the carrier density as determined from the Hall effect measurements, the validity of the Matthiessen rule and smooth evolution of NMR Knight shift, all suggesting the relation to localized electronic states rather than conduction electron system. Despite this, it strongly affects temperature dependent resistivity, with restoration of the $T-$linear resistivity above the transition. We further show that the resistivity above the transition is well described by Bloch-Gr\"uneisen function with notably higher Debye temperature of about 500~K as suggested in another heat capacity measurements \cite{Novikov} and with excess residual resistivity. We attribute this excess residual resistivity to excess contribution of rattling Ba atoms. Decay of this contribution below the transition is responsible for the anomalous temperature-dependent resistivity.

\section{Experimental}

All manipulations of starting materials were performed in an argon-atmosphere glovebox ($<$0.5 ppm O$_2$). Reagents were used as received: dendritic Ba pieces (Sigma Aldrich, 99.99\%), Cu powder (Alfa Aesar, 99.9\%), Ni powder (Alfa Aesar, 99.996\%), Si powder (Alfa Aesar, 99.99\%), red P powder (Alfa Aesar 98.9\%), Sn shot (Alfa Aesar, 99.80\%), and ultradry BaI$_2$ beads (Alfa Aesar, 99.999\%).

Single crystals of BaNi$_2$P$_4$ were grown using a Seed-Aided
Metal Flux method, which produced up to $\sim$6 mm$^3$-sized single
crystals and larger crystal agglomerates \cite{first}. For this technique,
the presynthesized single-phase powder of BaNi$_2$P$_4$ was sprinkled over
silica wool and kept inside the top, inverted Al$_2$O$_3$ crucible. The
reactants, Ba/Ni/P/Sn in 1:2:4:80 molar ratios using 30 g of Sn, were
loaded into the bottom Al$_2$O$_3$ crucible. The sealed ampule was
heated from room temperature to 1073 K at a rate of 30 K/h, then
kept at this temperature for 240 h. Afterwards, the ampule was
cooled down to 873 K over 40 h, annealed for 196 h at this
temperature, and then the furnace was turned off. The ampule was
tapped several times during the cooling and annealing processes to
drop the seed crystals into the bottom crucible and aid crystal growth.
After annealing, the Sn flux was dissolved by treating the sample in a
1:1 HCl/water solution.

Bulk powders of BaCu$_2$P$_4$ were prepared by loading Ba/Cu/P in a 1:2:4 molar ratio in a carbonized ampule. The ampule was evacuated, flame-sealed, and heated in a muffle furnace to 1098 K over 17 h, dwelled 140 h, then cooled to room temperature by turning off the furnace. The sample was annealed twice more following the same profile with intermittent grinding in the glovebox using an agate mortar and pestle. The final product was a black powder with trace (~2\%) admixture of Ba$_3$(PO$_4$)$_2$, and this powder was used for $^{31}$P NMR spectroscopy studies.
Single crystals of BaCu$_2$P$_4$ were obtained from a reaction containing Ba/Cu/Si/P/BaI$_2$ in a 8:16:1:30:2 molar ratio. The starting materials were flame-sealed in an evacuated carbonized silica ampule, then heated to 1133 K over 10 h, dwelled 144 h, and then naturally cooled to room temperature by turning off the furnace. The products were washed in DI water, filtered, then dried on air. Large (2-4 mm in length) shiny black rod-like crystals of BaCu$_2$P$_4$ were obtained and used for electrical resistivity and Raman spectroscopy measurements.

Samples for the electrical resistivity and Hall effect measurements were cut from larger crystals using a WS-22B high precision wire saw (from {\it Unipress}) and ground down to appropriate dimensions to remove tin flux inclusions.  Typical samples had dimensions on the order of (2-3 x 0.5 x 0.1) mm$^3$.  Contacts were made of 50 $\mu$m silver wire and were soldered with tin-silver alloy \cite{SUST,anisotropy}, resulting in a contact resistance below 100 $\mu \Omega$.    The temperature-dependent electrical resistivity measurements were performed in a {\it Quantum Design} PPMS (Physical Property Measurement System) setup in a temperature range of 2K-400 K using a four-probe configuration. The temperature range was extended up to 450~K using a Netzsch SBA 458 Nemesis system, with extended contacts to fit the size of the sample to the larger device. Measurements before and after irradiation were made on the same sample, thus excluding uncertainty in geometric factor determination.


For the Hall effect measurements, contacts were soldered to the side surfaces of samples in 5-probe configuration.   Measurements were preformed in a PPMS at constant temperatures, sweeping magnetic fields from 9 to -9 Tesla.   The antisymmetric part of the Hall resistance signal was then extracted and used for our analysis.

The electron irradiation was performed at the SIRIUS Pelletron facility of the Laboratoire des Solid\'es Irradi\'es at the \'Ecole Polytechnique in Palaiseau, France, using electrons with an energy of 2.5 MeV \cite{EMIR}.  The irradiation was done with the sample sitting in the liquid hydrogen ambient at around 20~K. This process introduces point-like defects throughout the entirety of the sample.  The samples were irradiated with a dose of 3.1 C/cm$^2$.  On warming up to room temperature, the defects partially anneal, with their density decreasing approximately by 30\%. Further annealing was proceeding during {\it in situ} resistivity measurements up to 400~K. Upon annealing, the number of defects in the crystals dropped to correspond to approximately a 1.7 C/cm$^2$ dose, as estimated by assuming linear resistivity variation with irradiated dose. 
We interchangeably use the terms unirradiated, as-grown, and pristine to describe samples before irradiation.


Polarized optical images were taken with a Leica DMLM polarization microscope \textit{Leica DMLM}.  The polarizer and analyzer were nearly crossed during imaging. The image shown in the left panel of Fig.~\ref{fig:Polarized_Optics}  was taken at room temperature and revealed a pattern of bright, nearly vertical stripes propagating through  growth steps. This is very similar to structural domains observed in iron-based superconductors below their tetragonal-to-orthorhombic transitions \cite{domains}. To confirm that the pattern is indeed due to domains, we heated the sample above the temperature of the tetragonal to orthorhombic transition $T_s$. A small soldering iron provided enough energy to heat the sample above the transition temperature (up to $\sim$450 K) without precision temperature control.  As expected, the domain pattern disappears above $T_s$.  Upon cooling, the domains reappeared, though their size depended on the rate of cooling.

\begin{figure} [b]
    \centering
    \includegraphics[width = 8.5cm]{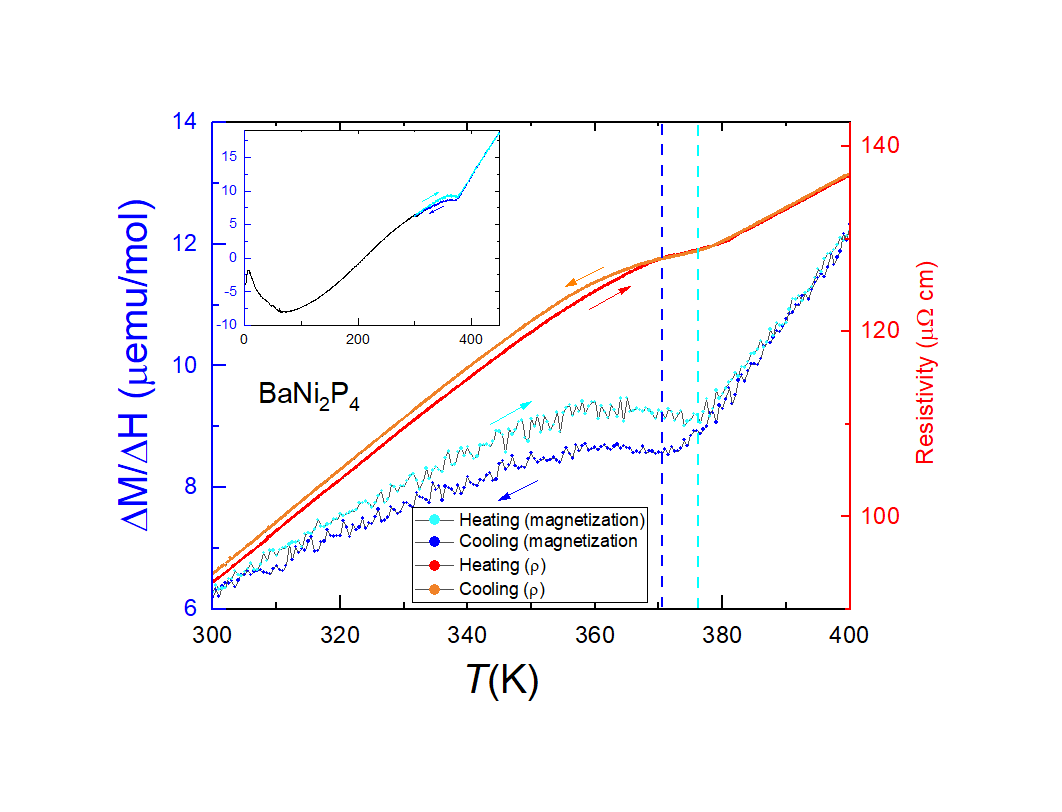}
    \caption{  
Temperature-dependent magnetic susceptibility of BaNi$_2$P$_4$
$\Delta M/ \Delta H$= ($M_1$-$M_2$)/ ($H_1$- $H_2$), where $H_1$ = 70
kOe, $H_2$ = 25 kOe (left scale, see \cite{first} for more details). Main panel shows zoom of the transition region measured on heating (cyan) and cooling (blue) in magnetization measurements revealing notable hysteresis. High temperature data were taken both on warming and on cooling
using 1 K/min heating/cooling rate. Red and orange lines are from resistivity measured on warming and cooling using the same protocol (right scale). Inset shows the magnetization data over the whole temperature range, including the low temperature data reported in Ref.~\onlinecite{first}. 
}
    \label{fig:magnetization}
\end{figure}

Magnetization measurements were performed using the VSM option of
the Quantum Design Magnetic Property Measurement System (MPMS3)
magnetometer. A fused silica sample holder was used for the 1.8 - 325 K range;
for 300 - 450 K measurements, the oven option was used with the sample mounted on
a heater stick with Zircar cement. For both mounts, the orientation of
the sample was kept the same. Whereas the fused silica sample holder has
virtually no background, small vertical shift of the high temperature
data was required to account for (temperature independent) contribution
of the heater stick and cement. Due to the presence of a small ferromagnetic
component, presumably from elemental Ni in the magnetic signal, the
Honda-Owen method, \cite{Honda,Owen}, $\chi = \Delta M/ \Delta H$, was used to
evaluate the intrinsic magnetic susceptibility (see \cite{first} for more detail).
Temperature-dependent magnetic susceptibility
clearly shows a structural transition at $T_s$, with the position depending on thermal sweep rate and direction. The  hysteresis between warming and cooling, $\sim$4 K, points to the first order
nature of the transition. The transition temperature is roughly consistent with
that reported in the literature \cite{Peierls} and that found in resistivity measurements. In the main panel of Fig.~\ref{fig:magnetization}, we directly compare the two measurements taken using an identical protocol for cooling and warming at a rate of 1 K/min. The exact position of the transition depends on the criteria used, and in the following we take $T_s=$376 $\pm$ 2K.

NMR measurements were performed on $^{31}$P ($I =$1/2, $\gamma _{\rm N}/2\pi$ = 17.2356 MHz/T) using a homemade phase-coherent spin-echo pulse spectrometer. Single crystals were crushed in an agate mortar to obtain powders to increase the signal to noise ratio in NMR measurements.  The $^{31}$P-NMR spectra were obtained by fast Fourier transform (FFT) of the NMR echo signals under an external magnetic field  $H$ of  $\sim$7.4089 T.   The $^{31}$P spin-lattice relaxation rate $(1/T_{1})$ was measured at the peak position of the spectra using a saturation recovery method.  $1/T_1$ at each temperature ($T$) was determined by fitting the nuclear magnetization $M_{\rm N}$ versus time $t$  using the exponential function $1-M_{\rm N}(t)/M_{\rm N}(\infty) = e^ {-t/T_{1}}$,  where $M_{\rm N}(t)$ and $M_{\rm N}(\infty)$ are the nuclear magnetization at time $t$ after saturation and the equilibrium nuclear magnetization at $t$ $\rightarrow$ $\infty$, respectively.   All the relaxation data measured in BaNi$_2$P$_4$ and BaCu$_2$P$_4$  were well fitted with this function.

Raman scattering measurements were conducted in a back-scattering geometry using 532 nm laser excitation at 5 mW in a crossed-polarization configuration. Data were obtained from single-crystal BaNi$_2$P$_4$ and BaCu$_2$P$_4$ samples, which were mounted in arbitrary orientations. Spectra at different temperatures have been normalized to the highest intensity phonon. Measurements performed above and below 300 K on two distinct BaNi$_2$P$_4$ crystals from the same growth batch revealed consistent Raman shifts. However, because the samples were mounted along arbitrary crystallographic directions relative to the incident polarization, variations in the relative intensities of the vibrational modes were observed between the two experimental runs.

 Electronic band structures of BaNi$_2$P$_4$ in LTO and HTT structures with spin-orbit coupling (SOC) in density functional theory \cite{DFT1,DFT2} (DFT) have been calculated using the Perdew-Burke-Ernzerhof (PBE) \cite{DFT3} exchange-correlation functional, a plane-wave basis set, and the projected augmented wave method \cite{DFT4} as implemented in VASP \cite{DFT5,DFT6}. For accurate density of states (DOS) calculations, the tetrahedron method \cite{DFT7} with a Monkhorst-Pack \cite{DFT8} (16$\times$16$\times$16) $k$-point mesh, including the $\Gamma$ point and a kinetic energy cutoff of 269.5 eV, has been used. An atomic radius of 1.6 \AA~ has been used to calculate the density of states (DOS) at the different P sites in the LTO structure. For Hall coefficient and transport calculations, the semiclassical Boltzmann formula with the constant scattering rate approximation has been used \cite{DFT9}.


Inelastic neutron scattering (INS) measurements of the phonon density of states of BaNi$_2$P$_4$ were performed using the ARCS time-of-flight spectrometer at the Spallation Neutron Source at Oak Ridge National Laboratory. Polycrystalline samples were encased in an aluminum can, and data were collected over the temperature range of 25 K – 600 K using an incident energy of 90 meV. Empty can data were measured under identical conditions and subtracted from all raw data. Data reduction was performed using the Mantid package,\cite{Mantid} and the neutron-weighted phonon density of states was extracted using a standard procedure.\cite{INS}

\section{Results}

\subsection{Transport measurements}

\begin{figure}
    \centering
    \includegraphics[width = \linewidth]{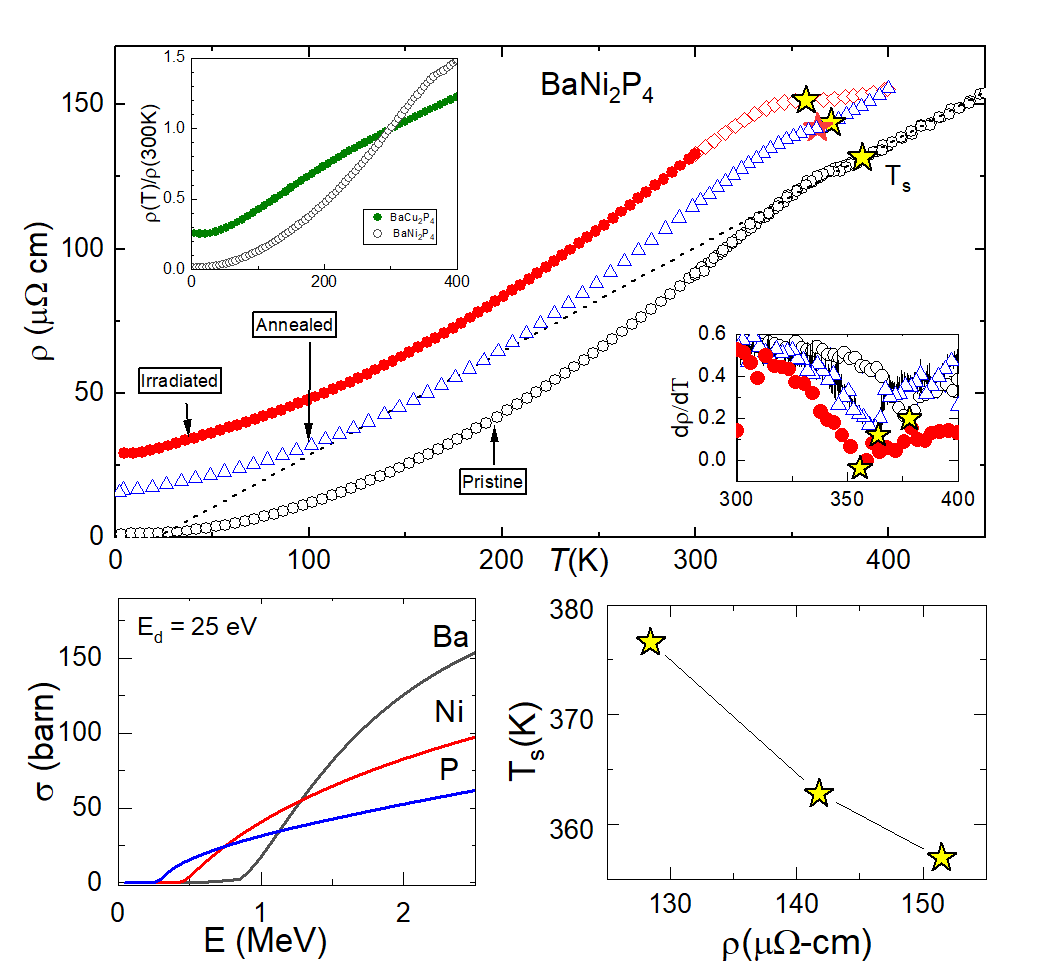}
    \caption{Top panel. Variation of the temperature-dependent resistivity, $\rho (T)$, of \BaNiP~ with controlled disorder.  Black symbols represent the data in pristine state, red solid dots are after 3.1 C/cm$^2$. When the temperature of irradiated sample rises above 300~K, sample annealing starts (open red diamonds). The resistivity transients to a new value depending on the highest temperature reached. The $\rho(T)$ after 400~K annealing is stable (open blue up-triangles). The black dotted line is linear fit through the $\rho(T)$  above the structural transition in the pristine state. Both linear behavior of resistivity and extrapolation to positive temperatures are in line with expectations for a good metal, see Fig.~\ref{fig:DebyeTemp}. Nearly parallel shift with respect to the pristine sample is observed after annealing. The bottom right inset in the top panel shows zoom of the temperature-dependent resistivity derivative in the transition area with stars highlighting the $T_s$ position (reproduced in the main panel) and its shift in response to disorder.  The left  top inset in the top panel shows a comparison between the resistivity behavior of BaNi$_2$P$_4$ (black open circles) and BaCu$_2$P$_4$ (green solid circles). Bottom left panel shows 
defects creation cross-sections for Ba (top black curve), Ni (red curve) and P (blue curve) in BaNi$_2$P$_4$ 
as function of electron energy assuming the displacement energy threshold,
$E_d=25~\mathrm{eV}$.  Bottom right panel shows variation of the structural transition temperature with resistivity of the sample immediately above the transition, used as a proxy for sample disorder level.}
    \label{fig:Irradiation}
\end{figure}

In the top panel of Fig.~\ref{fig:Irradiation}, we show the temperature-dependent resistivity of a pristine sample of \BaNiP~ (black curve) over the temperature range that includes the tetragonal-orthorhombic transition at around 378~K in these measurements. The transition (indicated with the star) leads to a feature in the $\rho(T)$ most clearly seen in the temperature-dependent resistivity derivative $d \rho(T)/d T$ (right bottom inset in the top panel). The resistivity above the transition, though in a very limited range in these measurements, is close to linear, with the dotted black line showing a linear fit through the data. The red curve shows $\rho(T)$ of the same sample after electron irradiation with 3.1 C/cm$^2$. Measurements of the irradiated sample were initially performed by cooling the sample from 300~K down to 2~K and then warming it up to 400~K. As can be seen, the curves are completely reproducible (red solid dots) below 300~K. Upon reaching 300~K, the sample resistivity starts to depend on defect annealing, determined by the highest temperature reached (red open circles). As a result, nearly parallel resistivity shifts between pristine and irradiated curves start to decrease, and a new shift corresponds to a lower defect density. Upon reaching 400~K, the new run (blue up-triangles) was taken down to 2~K and up to 400~K, revealing no further change in resistivity and complete reproducibility between cooling and warming cycles. 
The resistivity curve shows a nearly parallel shift compared with the curve for the pristine sample, suggesting the validity of the Matthiessen rule.  This partial annealing reduces the residual resistivity by about a factor of 2. Assuming a linear relation of residual resistivity with the irradiation dose, verified experimentally in other systems obeying the Matthiessen rule \cite{PRX,PdTe2}, this partially annealed state corresponds to an effective dose of 1.7~C/cm$^2$. Note that the transition temperature is visibly suppressed by disorder, from 378~K to 355~K (bottom right panel in Fig.~\ref{fig:Irradiation}).

The rate of suppression $\Delta T_s/T_{s0}=$2\% per 1 C/cm$^2$ 
in BaNi$_2$P$_4$ can be compared to that in the other systems.  In charge density wave 
materials (Sr,Ca)$_3$(Rh,Ir)$_4$Sn$_{13}$ \cite{Elizabeth3413} the rate of suppression is 
about the same, $\sim$ 2\% per 1 C/cm$^2$. In magnetite F$_3$O$_4$  the rate of suppression 
is somewhat smaller, about 0.5\% per 1 C/cm$^2$ \cite{magnetite}. Of note, Jahn-Teller mechanism of the transition is also invoked for the explanation  of the Verwey transition here.

\begin{figure}
    \centering
    \includegraphics[width = \linewidth]{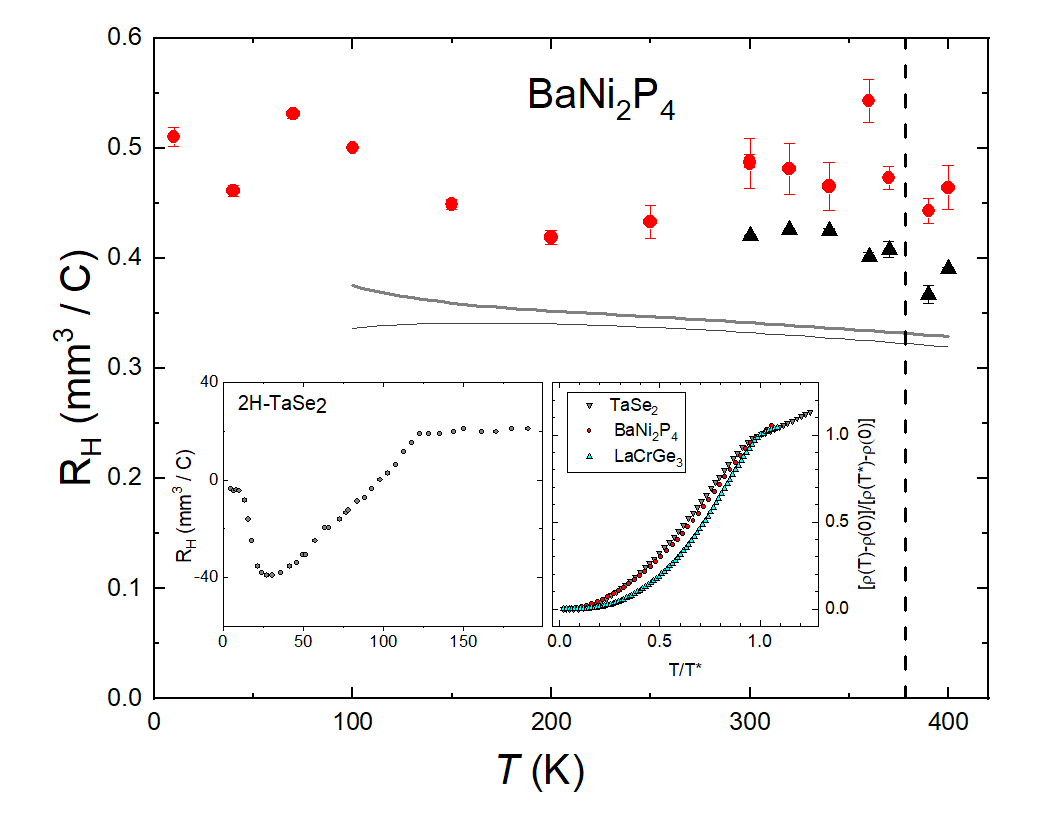}
    \caption{Temperature-dependent Hall effect measured in two samples of BaNi$_2$P$_4$ (red solid circles and black solid up-triangles), no change of the Hall constant on crossing $T_s$ (vertical dashed line) beyond approximately 10\% uncertainty of the Hall effect measurements.   Thicker and thinner gray lines show Hall effect in the tetragonal and orthorhombic phases from the calculated band structure assuming constant scattering rate. The effect of the transition at $T_s$ is negligible and well below error bars of the measurements. Right inset shows comparisons of the temperature dependent resistivity of BaNi$_2$P$_4$ (red curve), of the anomalous charge density wave material TaSe$_2$ (black curve) (from Ref.\onlinecite{Naito1982}) and of ferromagnetic LaCrGe$_3$ (green symbols, after \onlinecite{LaCrGe3}), using normalized resistivity, $\rho/\rho(T^*)$ and temperature $T/T^*$ scales. Despite close resemblance of the resistivity curves for BaNi$_2$P$_4$ and TaSe$_2$, the two compounds find quite different temperature dependent Hall effect (shown for TaSe$_2$ in the left inset).}
    \label{fig:Hall}
\end{figure}

To get an insight into a notable change of the temperature dependent resistivity below $T_s$ from a $T-$linear above to a strongly non-linear below, we need to understand the change in the carrier density over the transition. Indeed, the temperature-dependent carrier density, as exemplified by two-dimensional charge density wave material TaSe$_2$, can bring notable non-linearity to the $\rho (T)$ \cite{Naito1982}. To check this hypothesis, we performed Hall effect measurements in BaNi$_2$P$_4$ through the transition temperature, as shown in Fig.~\ref{fig:Hall}. We found the Hall constant $R_H=\rho_{xy}d/B$, where $\rho_{xy}$ is Hall resistance, $d$ is sample thickness, and $B$ is magnetic field, to be independent of temperature within the noise of the data, both above and below the transition. The experimental results are in semi-quantitative agreement with the magnitude of the Hall effect calculated using the semi-classical Boltzmann transport theory and the constant scattering rate approximation \cite{DFT9,Chaput} with DFT-calculated band structure. The constant scattering rate approximation is good only for high temperatures, so the calculations are shown above 100~K  
both for the tetragonal phase (solid line) and the orthorhombic phase (dashes) in Fig.~\ref{fig:Hall}.  
This nearly temperature independence suggests that the variation of the conduction electron density plays a minor role in the transition. For reference, we compare the temperature dependent resistivity of TaSe$_2$ and BaNi$_2$P$_4$ in the bottom right inset using normalized resistivity, $\rho/\rho(T_s)$, and temperature $T/T_s$ scales. Despite the close resemblance of the resistivity curves, the two compounds exhibit quite different temperature dependent Hall effects (shown for TaSe$_2$ in the left inset in Fig.~\ref{fig:Hall}. ). We also compare the $\rho(T)$ of BaNi$_2$P$_4$ with the case of dominant magnetic scattering in the itinerant ferromagnet LaCrGe$_3$ \cite{LaCrGe3}. Similar to BaNi$_2$P$_4$, Matthiessen rule is obeyed upon irradiation in the latter material, suggesting negligible transformation of the electronic structure, but a close similarity of $\rho(T)$ due to a loss of spin-disorder scattering in the ferromagnetically ordered state.

\subsection{NMR}

\begin{figure}[b]
 \centering
 \includegraphics[width = \linewidth]{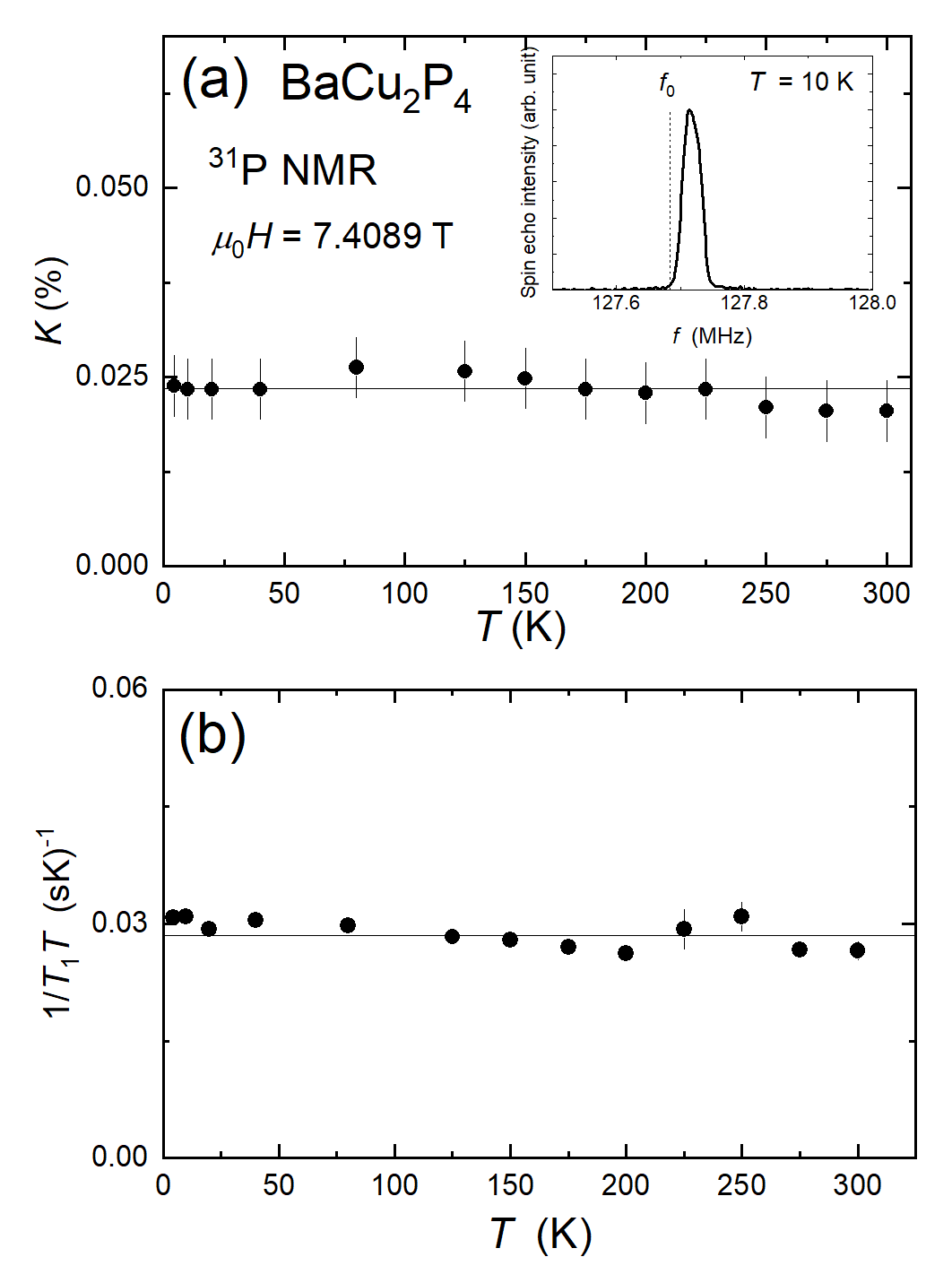} 
\caption{(a) Temperature dependence of NMR shift $K$ for BaCu$_2$P$_4$. The inset shows the $^{31}$P-NMR spectrum at $T$ = 10 K under a magnetic field of $\mu_0H$ = 7.4089 T.   The vertical dashed line in the inset represents the corresponding zero-shift positions in the $^{31}$P NMR spectrum (Larmor frequency $f_0$). (b) Temperature dependence of the $^{31}$P spin-lattice relaxation rate divided by temperature (1/$T_1T$) in BaCu$_2$P$_4$. The horizontal solid lines in (a) and (b) are guides for the eye. }
\label{fig:Fig1_NMR}
\end{figure}

NMR measurements using the naturally abundant $^{31}$P nucleus as a probe provide additional insight into the changes in the electronic band structure of the materials. We performed comparative studies using BaCu$_2$P$_4$ as a reference compound to understand the anomalous behavior of BaNi$_2$P$_4$.  The inset of Fig.~\ref{fig:Fig1_NMR} (a) shows the $^{31}$P NMR spectrum measured at $T$ = 10 K under a magnetic field of 7.4089 T in BaCu$_2$P$_4$. Since the $^{31}$P nucleus has $I$ = 1/2, a single line was observed as expected. The observed spectra are nearly independent of temperature, with a full width at half maximum (FWHM)  of  $\sim$33 kHz ($\sim$19 Oe). The NMR shift ($K$), determined by the peak position of the spectra, also shows temperature independent behavior as in  Fig.~\ref{fig:Fig1_NMR} (a).  Fig.~\ref{fig:Fig1_NMR} (b) shows the temperature dependence of $^{31}$P spin-lattice relaxation rate divided by temperature (1/$T_1T$), which exhibits a  1/$T_1T \approx$ constant behavior.  The temperature independent behavior for both $K$  and 1/$T_1T$ is expected for a normal metal, similar to the temperature independent Hall effect and magnetic susceptibility. 

\begin{figure}[b]
\centering
\includegraphics[width = \linewidth]{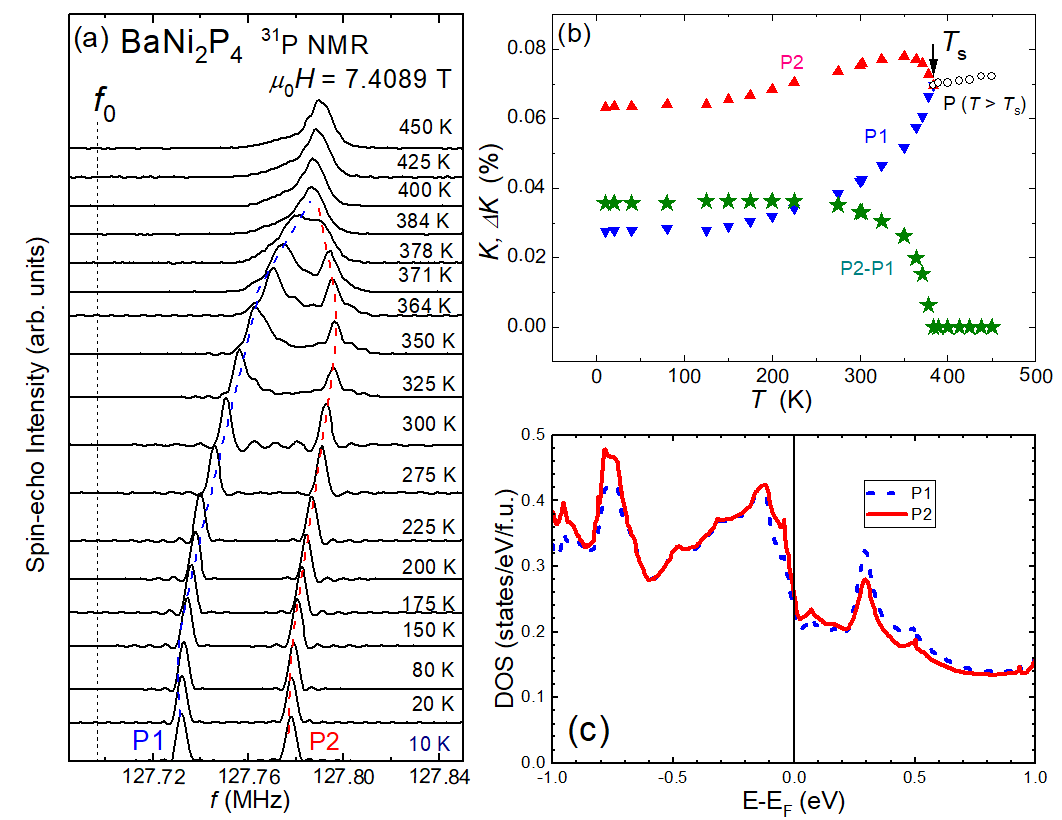} 
\caption{(a) Temperature evolution of $^{31}$P-NMR spectra in BaNi$_2$P$_4$ under a magnetic field 7.4089 T. Below $T_{\rm s}$, the spectra split into two lines (P1 and P2).The vertical dashed line represents the corresponding zero-shift position in the $^{31}$P-NMR spectrum (Larmor frequency $f_0$). (b) Temperature dependence of NMR shift $K$ for P1 [$K(P1)$: blue down-pointing triangles] and P2 [$K$(P2): red up-pointing triangles] below $T_{\rm s}$ and for the peak (open circles) above $T_{\rm s}$. 
Green stars show temperature dependence of the NMR line splitting $\Delta$$K$ [= $K$(P2)$-$$K$(P1)].
(c) Calculated density of states projected for P1 and P2. }
\label{fig:Fig2_NMR}
\end{figure}

In stark contrast, the NMR spectra of BaNi$_2$P$_4$ reveal strong temperature dependence as shown in Fig.~\ref{fig:Fig2_NMR}(a). Although a single line is observed above $T_{\rm s}$ = 378 K as in the case of BaCu$_2$P$_4$, the line splits into two lines (P1 and P2) with decreasing temperature below $T_{\rm s}$. This is due to the structural phase transition from the HTT to the LTO phases where two crystallographically inequivalent P sites exist \cite{Peierls} while only one P site exists in the HTT phase \cite{Peierls}. 
 
Figure~\ref{fig:Fig2_NMR}(b) shows the temperature dependence of NMR shifts $K$ for P1 [$K(P1)$] and P2 [$K$(P2)] below $T_{\rm s}$ together with $K$ for the single peak observed  above $T_{\rm s}$.
Just below $T_{\rm s}$, $K$(P2) slightly increases, then decreases with a local maximum around  350 K and levels off at low temperatures below $\sim$125 K. On the other hand, $K$(P1) gradually decreases with decreasing temperature below $T_{\rm s}$ and becomes temperature independent below $\sim$125 K.
The NMR shift generally has two contributions: the electron spin part $K_{\rm spin}$ and the temperature independent part $K_0$: $K$ = $K_{\rm spin}$ + $K_0$, where $K_0$ originates from the core electrons of the P atoms and their local environment as well.
The different values of $K$ suggest that $K_{\rm spin}$ and $K_0$ can differ for the two P sites. 
As $K_{\rm spin}$ is proportional to the local density of states (DOS) at the Fermi Energy $\cal{D(E_{\rm F}}$), the different values of $K$(P1) and $K$(P2) suggest that $\cal{D(E_{\rm F}}$) for P2 is greater than that of P1. Also, as shown below, the results of 1/$T_1T$ show the different DOS at P1 and P2. 
It is noted that the change in the bond distance between Ni and P atoms can lead to a change in $K_0$. Actually, we found different values of $K_0$ for the P1 and P2 sites, as shown below. 

As the temperature dependence of $K$(P2) and $K$(P1) near the $T_{\rm s}$ reflects the nature of the structural phase transition, we plot  $\Delta$$K$ = $K$(P2)$-$$K$(P1)  in  Fig.~\ref{fig:Fig2_NMR} (b) (green stars). As can be seen, $\Delta$$K$ increases rapidly below $T_{\rm s}$ and becomes constant below $\sim$225 K.

To check the difference in DOS at the P sites, we have performed DFT calculations in the LTO phase, where two distinct bond lengths between P and Ni ions exist, making two crystallographically inequivalent P sites [P(1) and P(2)] \cite{Peierls}. With a set of lattice parameters of  $a$ = 6.620 \AA, $b$ = 6.470 \AA, and $c$ = 5.785 \AA~in the LTO phase ($Immm$) \cite{Peierls}, the bond length between Ni-P(1) is 2.218 \AA~in the $ac$ plane which is shorter than 2.240 \AA~for the Ni-P(2) bond-length in the $bc$ plane \cite{Peierls}. The calculated DOS near the Fermi level at P(2) is found to be slightly greater than that of P(1) as shown in the bottom panel (c) of  Fig.~\ref{fig:Fig2_NMR}(c). The different DOS between the two P sites is consistent with the NMR data, and thus, we can assign P1 and P2 defined in the NMR spectrum measurements to P(1) and P(2), respectively.

\begin{figure}[b]
\centering
\includegraphics[width = 0.95\linewidth]{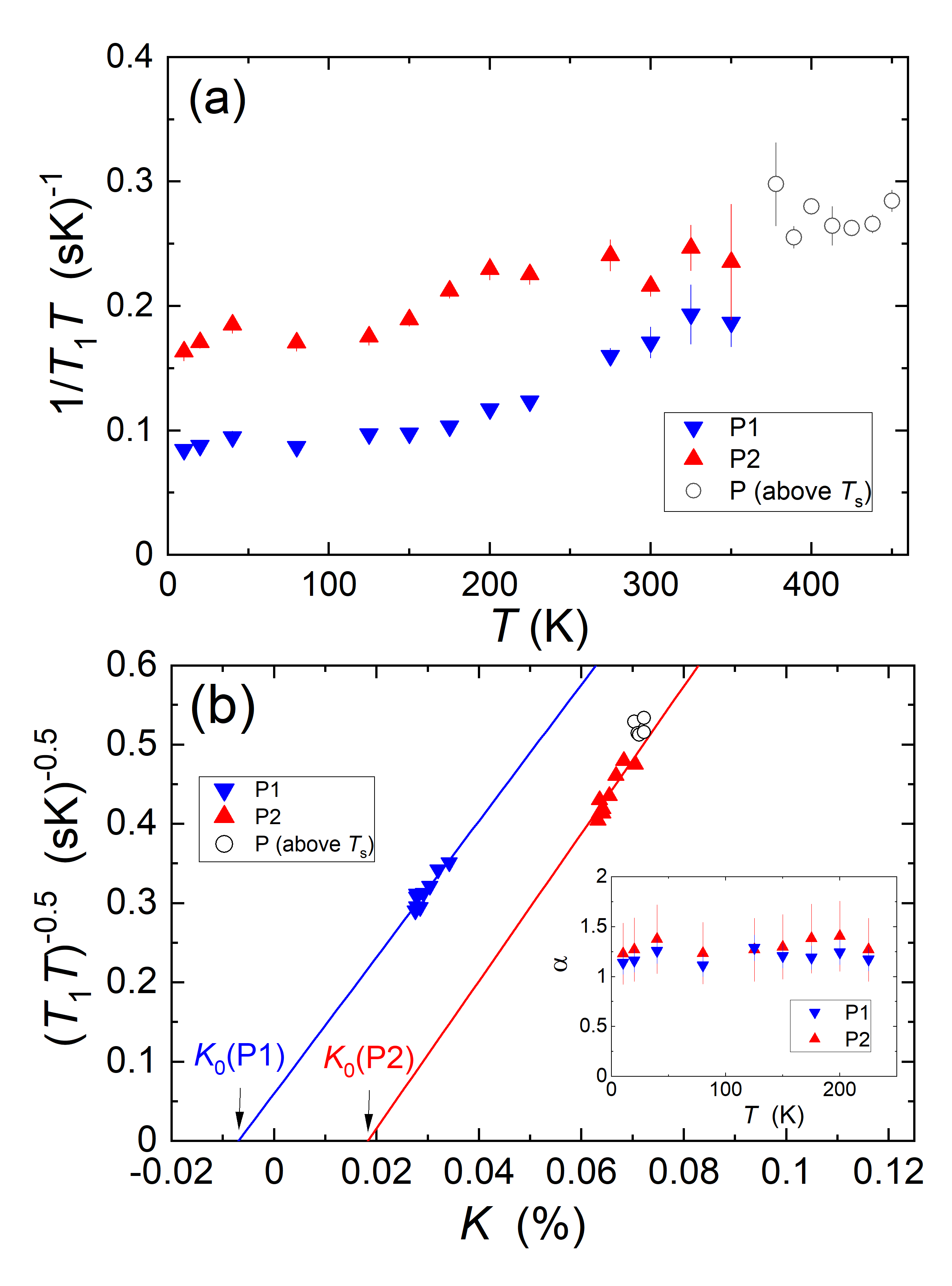} 
\caption{(a) Temperature  dependence of the $^{31}$P spin-lattice relaxation rate divided by temperature (1/$T_1T$) for P1 (solid blue down-pointing triangles) and P2 (solid red up-pointing triangles) below $T_{\rm s}$ and for the P site (open circles) above $T_{\rm s}$ of BaNi$_2$P$_4$. (b) (1/$T_1T$)$^{1/2}$ versus $K$ plot for P1 (blue down-pointing triangles) and P2 (red up-pointing triangles) together with data (open circles) above $T_{\rm s}$. The inset shows the temperature dependence of $\alpha$ for P1 and P2. }
\label{fig:Fig3_NMR}
\end{figure}

To get further information about local physical properties at the P sites in BaNi$_2$P$_4$, we also measured $^{31}$P-1/$T_1T$ with the results shown in Fig.~\ref{fig:Fig3_NMR}(a).
Above $T_{\rm s}$, 1/$T_1T$ is nearly independent of temperature within our experimental uncertainty, consistent with a metallic state.
Below $T_{\rm s}$, $1/T_1T$ gradually decreases with decreasing temperature and exhibits a $1/T_1T$ constant behavior below $\sim$150 K; again, the 1/$T_1T$ = constant is expected for metallic states. 
It is noted that the value of 1/$T_1T$ for P2 is greater than that of P1. Since 1/$T_1T$ is proportional to the squared $\cal D(E_{\rm F})$,   these results indicate that the local DOS at P2 is higher than that of P1. This is consistent with the results of the NMR shift described above.

Now we discuss the magnetic fluctuations in BaNi$_2$P$_4$ based on  a modified Korringa relation analysis using the $T_1$ and the electron spin part of the NMR shift ($K_{\rm spin}$) data.
Within a Fermi liquid picture, as described above, $1/T_1T$ is proportional to the square of the DOS near the Fermi energy ${\cal D}(E_{\rm F})$ and $K_{\text{spin}}$ is proportional to ${\cal D}(E_{\rm F})$. 
In particular, $T_1TK_{\text{spin}}^2$  = $\frac{\hbar}{4\pi k_{\rm B}} \left(\frac{\gamma_{\rm e}}{\gamma_{\rm N}}\right)^2$ = $S$, which is the Korringa relation.  
Deviations from $S$ can reveal information about electron correlations in materials \cite{Moriya1963,Narath1968}, which are expressed via the parameter $\alpha=S/(T_1TK_{\text{spin}}^2)$. 
For instance, the enhancement of $\chi(\mathbf{q}\neq 0)$ increases $1/T_1T$ but has little or no effect on $K_{\text{spin}}$, which probes only the uniform $\chi$ with $\mathbf{q}$ = 0.  
   Thus  $\alpha >1$ for AFM correlations and $\alpha <1$ for FM correlations.

To estimate $\alpha$ value, one needs to obtain $K_{\rm spin}$ from the experimentally observed $K$.
From the relationship between $1/T_1T$ and $K$ described above, $1/\sqrt{T_1T}$ is proportional to $K$ =  $K_{\rm spin} + K_0$.
Thus by plotting $1/\sqrt{T_1T}$ versus $K$  with $T$ as an implicit parameter, one can obtain $K_0$ and thus $K_{\rm spin}$.
In fact, as shown in Fig.~\ref{fig:Fig3_NMR}(b),  a linear relation between  $1/\sqrt{T_1T}$ and  $K$ =  $K_{\rm spin} + K_0$ can be observed, where we plotted the data below 225 K to avoid the effects on the $K$ and $1/T_1T$ data from the structural phase transition.
From the horizontal intercept, $K_0$s are estimated to be  -0.007\% and 0.018\% for P1 and P2, respectively.
Utilizing the values of $K_0$, we estimated $\alpha$ values whose temperature dependence is shown in the inset of    Fig.~\ref{fig:Fig3_NMR}(b). Although experimental uncertainty is relatively large, $\alpha$ seems to be close to unity for both P sites. This would suggest no strong magnetic fluctuations (or electronic correlations) in the metallic state below $T_{\rm s}$ in BaNi$_2$P$_4$.
It is noted that, above $T_{\rm s}$, we were not able to estimate $K_0$ since 1/$T_1T$ is nearly independent of temperature, which makes the analysis difficult (see the open circles in Fig.~\ref{fig:Fig3_NMR}(b)).  Nevertheless, as the open circles in the figure seem to be close to the line fit to the P2 data, this may suggest no significant magnetic fluctuations even in the HTT phase.

\subsection{Raman scattering}

\begin{figure}
    \centering
    \includegraphics[width = \linewidth]{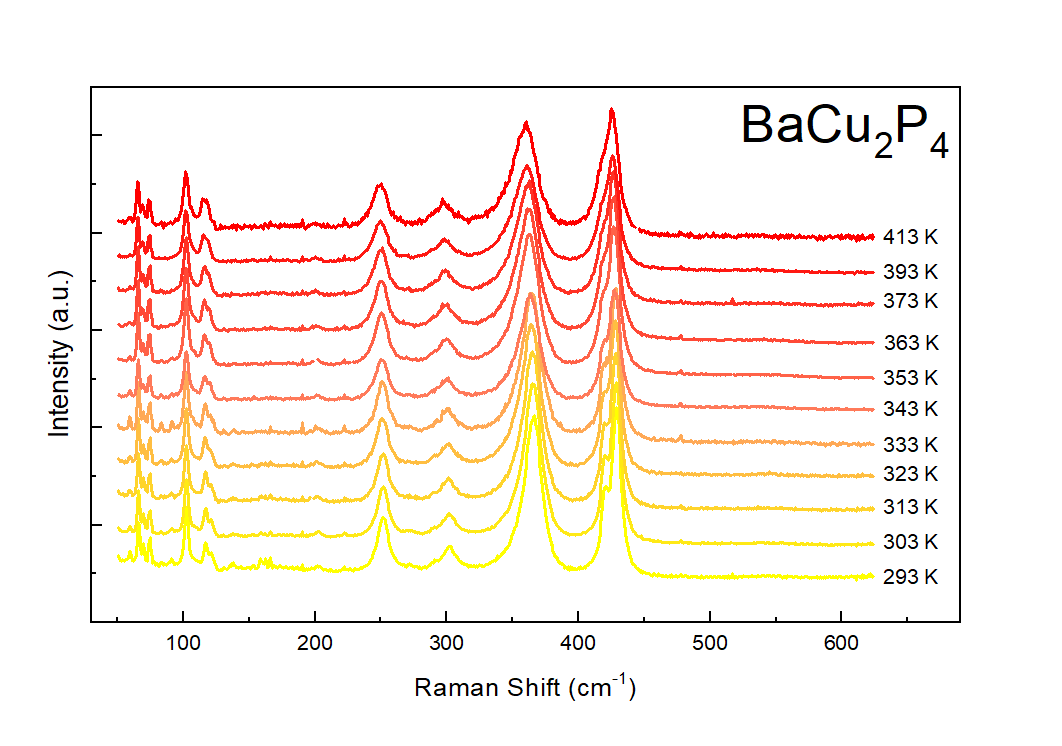}
    \caption{Temperature evolution of the Raman spectra in BaCu$_2$P$_4$ at temperatures above 300~K, showing negligible evolution of the spectra. }
    \label{fig:RamanHTBaCu2P4}
\end{figure}

\begin{figure}
    \centering
    \includegraphics[width = \linewidth]{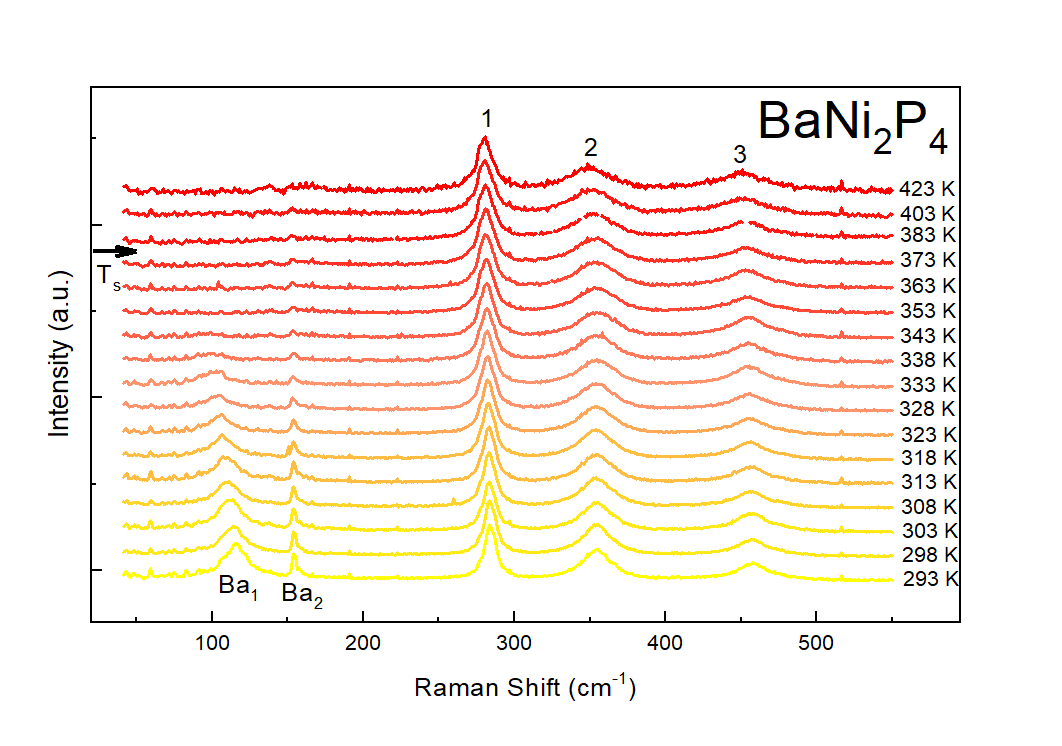}
    \caption{Temperature evolution of the Raman spectra in BaNi$_2$P$_4$ at temperatures above 300~K. Note complete suppression of Ba$_1$ and Ba$_2$ lines in the tetragonal phase above 373~K. }
    \label{fig:RamanvsT}
\end{figure}

\begin{figure}
    \centering
    \includegraphics[width = \linewidth]{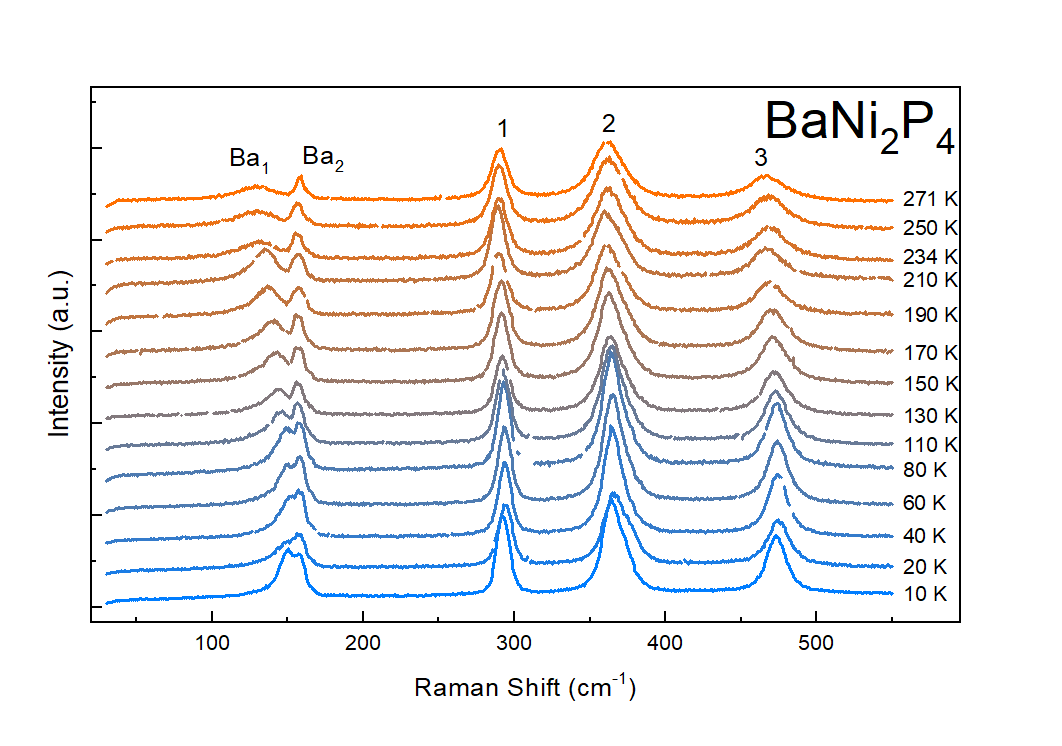}
    \caption{Temperature evolution of the Raman spectra in BaNi$_2$P$_4$ at temperatures below  300~K. }
    \label{fig:RamanHT}
\end{figure}


\begin{figure}
    \centering
    \includegraphics[width = \linewidth]{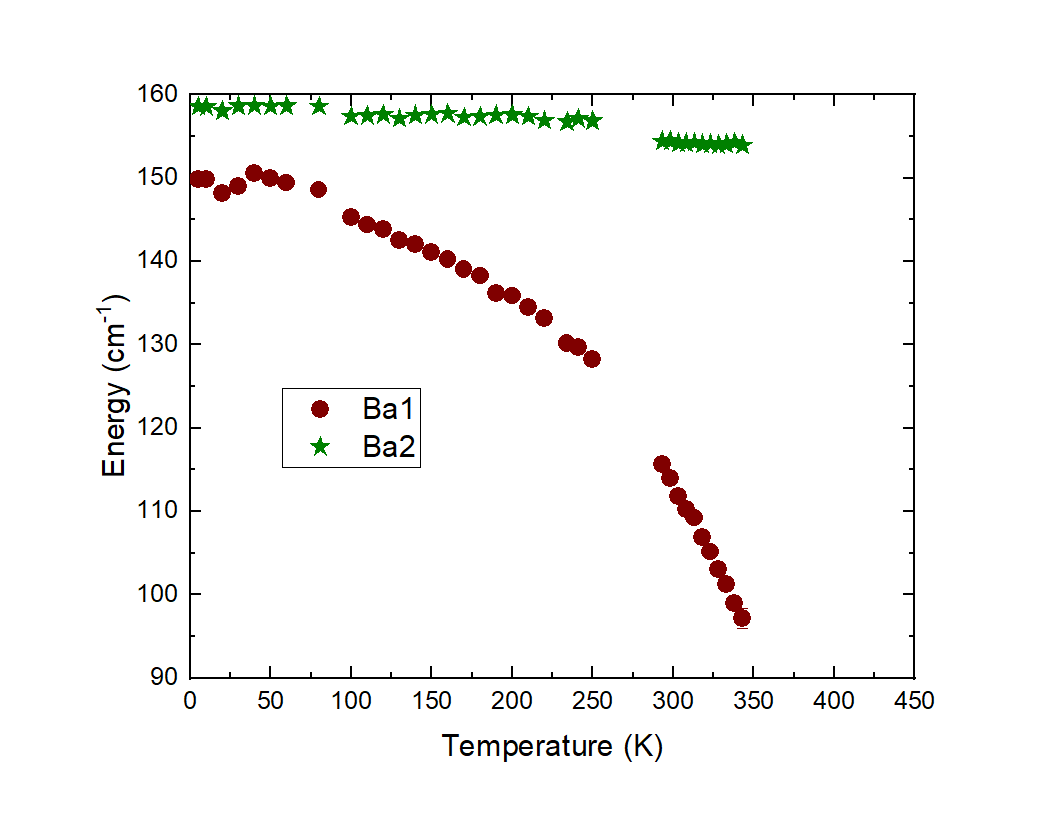}
    \caption{  Temperature dependence of the Ba$_1$ (brown circles) and Ba$_2$ (green stars) line positions in the Raman spectra.   
 }
    \label{fig:Ba1Ba2}
\end{figure}

In Fig.~\ref{fig:RamanHTBaCu2P4}, we show the evolution with temperature  of the Raman spectra in the reference compound BaCu$_2$P$_4$ at temperatures above 300~K. The spectra show some broadening on heating to 413~K, but no new lines appear and no significant transformations in the spectra are visible. Most notably, the group of lines between 50 and 150 cm$^{-1}$, due to Ba rattling inside the cage \cite{first} remains unchanged. 

Fig.~\ref{fig:RamanvsT} and Fig.~\ref{fig:RamanHT} show the spectra of Raman scattering in BaNi$_2$P$_4$ below and above room temperature, respectively. While the temperature dependence of the position and the width of the peaks denoted as  Ba$_1$ and Ba$_2$ becomes notable for temperatures as low as 150~K,  it becomes very pronounced when approaching $T_s$ from below. Ba$_1$ line softens and broadens and eventually becomes indistinguishable  for temperatures above 353~K. The Ba$_2$ mode disappears at T$_s$ without significant softening, Fig.~\ref{fig:Ba1Ba2}. This behavior is consistent with a structural transition to a higher-symmetry phase where this particular vibration becomes Raman-inactive due to selection rules.
The high frequency modes 1, 2, and 3 (shifts above 250 cm$^{-1}$) do not reveal any changes, as illustrated in Fig.~\ref{fig:RamanvsT} and Fig.~\ref{fig:RamanHT}.

This difference in the behavior of low frequency modes in Ni and Cu compounds is instrumental in understanding the physics of the phase transition. It is clear that Ba$_1$ line in BaNi$-2$P$_4$ represents  the soft mode of the transition \cite{softmodes}. Similar to the soft mode observed, for example, in SrTiO$_3$, the mode is related to the distortion of the cage in the orthorhombic phase, not to the existence of the cage per se. The topology of M$_8$P$_1$$_6$ cages is different in BaNi$_2$P$_4$ and BaCu$_2$P$_4$ \cite{SrNi2P4} and the difference in the temperature evolution of the lines suggests that the motion of Ba atom remains restricted inside the cage in the Cu compound so that the vibration frequencies do not change. On the contrary, softening on Ba$_1$ line and loss of amplitude of Ba$_2$ line suggest that Ba atom is not bound to the cage atoms in the HTT phase of the BaNi$_2$P$_4$. Distortion of the cage below $T_s$ is related to a shift of the Ba ion from the central position in HTT phase, which naturally leads to the existence of P ions closer to and further away from the central ion and a difference in the local environment, which may lead to different values in $K_0$ for P(1) and P(2) in the NMR spectra. Thus, the observed different values of $K_0$ may support the scenario.

In Fig.~\ref{fig:Ba1Ba2} we plot the energy of the Ba$_1$ and Ba$_2$ vibrations in the Raman spectra. The energy of the Ba$_1$ vibration is indirectly related to the order parameter of the structural phase transition (that is, orthorhombic distortion), similar to NMR line splitting, Fig.~\ref{fig:Fig2_NMR}.(b) 
Comparison of the temperature dependence of NMR splitting and of the Raman line position reveals one interesting difference. NMR line splitting shows a notably faster increase below the transition and clear saturation below 200~K. The NMR shift includes both chemical shift and Knight shift contributions; the latter is supposed to be temperature independent in a metal. This suggests that the distortion of the cage mainly happens right below the transition, and it plays a very important role in the overall NMR shift. 

\subsection{Inelastic neutron scattering}

\begin{figure}
    \centering
    \includegraphics[width = \linewidth]{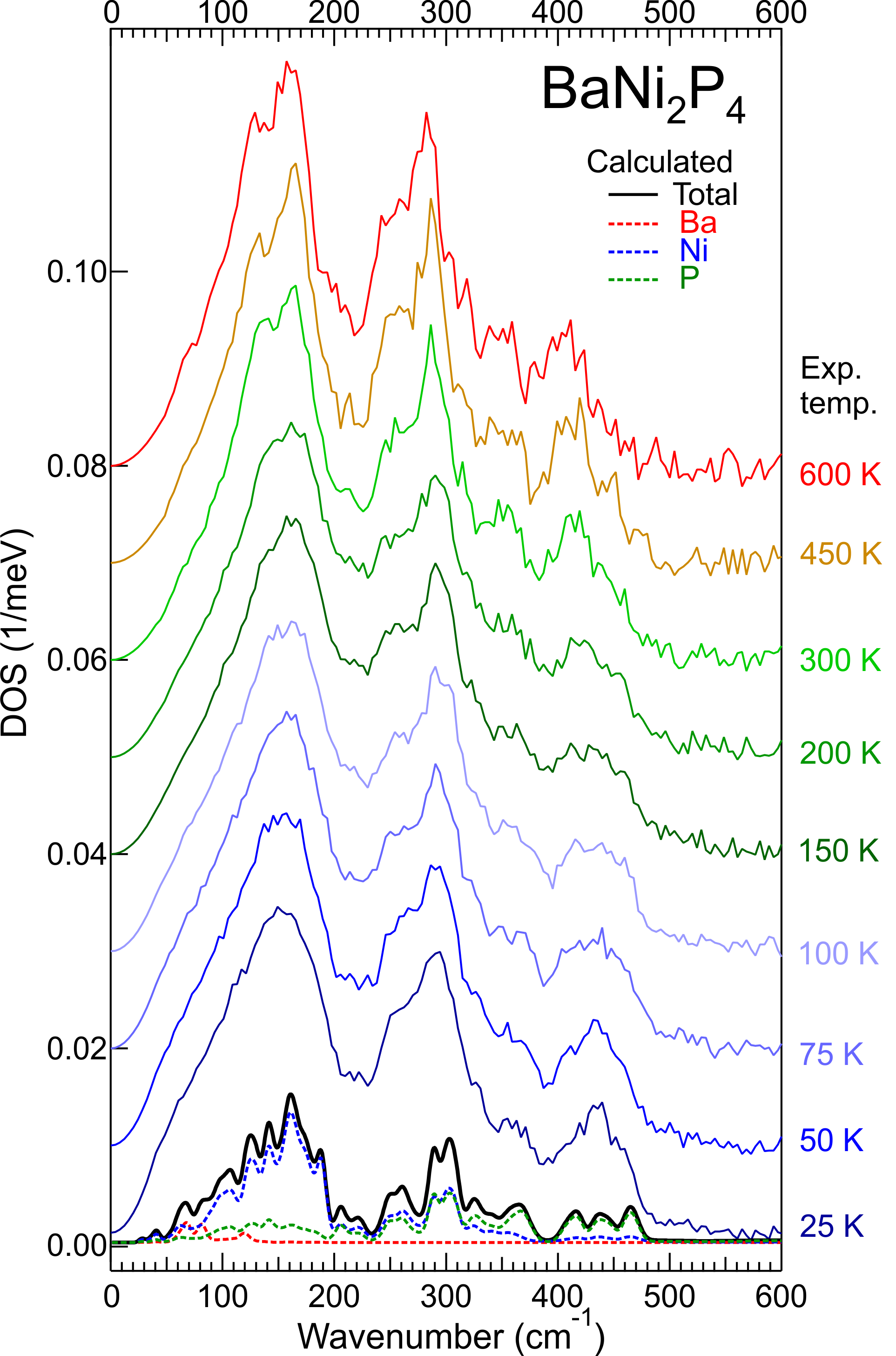}
    \caption{  Neutron-weighted phonon density of states extracted from
inelastic neutron scattering data as a function of temperature. Curves
are offset for clarity. The bottom plot is a calculation weighted by
neutron cross-section PDOS for BaNi$_2$P$_4$.    
 }
    \label{fig:INS}
\end{figure}

Inelastic neutron scattering measurements reported in our initial report below room temperature \cite{first} were extended to higher temperatures, up to 600~K, to cover the transition range at $T_s$. As reported previously, partial densities of states (PDOS) are consistent with simulations, as shown by the bottom curves in Fig.~\ref{fig:INS}. The low-energy 
portion of the measured PDOS shows a stiffening with increasing
temperature and non-Debye behavior for temperatures less
than 100 K, possibly reflecting the anharmonicity of Einstein-like modes related to uncorrelated Ba ionic motions within the Ni-P framework below 100~K. The rest of the PDOS shows little change upon warming, indicating that the Ni$-$P framework
is more harmonic and less sensitive to thermal changes. 

Ba vibrational
amplitude dominates the flat regions at the top of acoustic
branches as well as some low-energy portions of optical
branches, which supports the notion of low-energy, large amplitude
Ba vibrations. However, since the Ba modes are
clearly hybridized with acoustic modes, there is not a distinct
decoupling into Ba modes versus framework modes. BaNi$_2$P$_4$
exhibits a splitting of the Ba position, resulting in short Ba framework
interactions.

\section{Discussion}

\begin{figure}
    \centering
    \includegraphics[width = \linewidth]{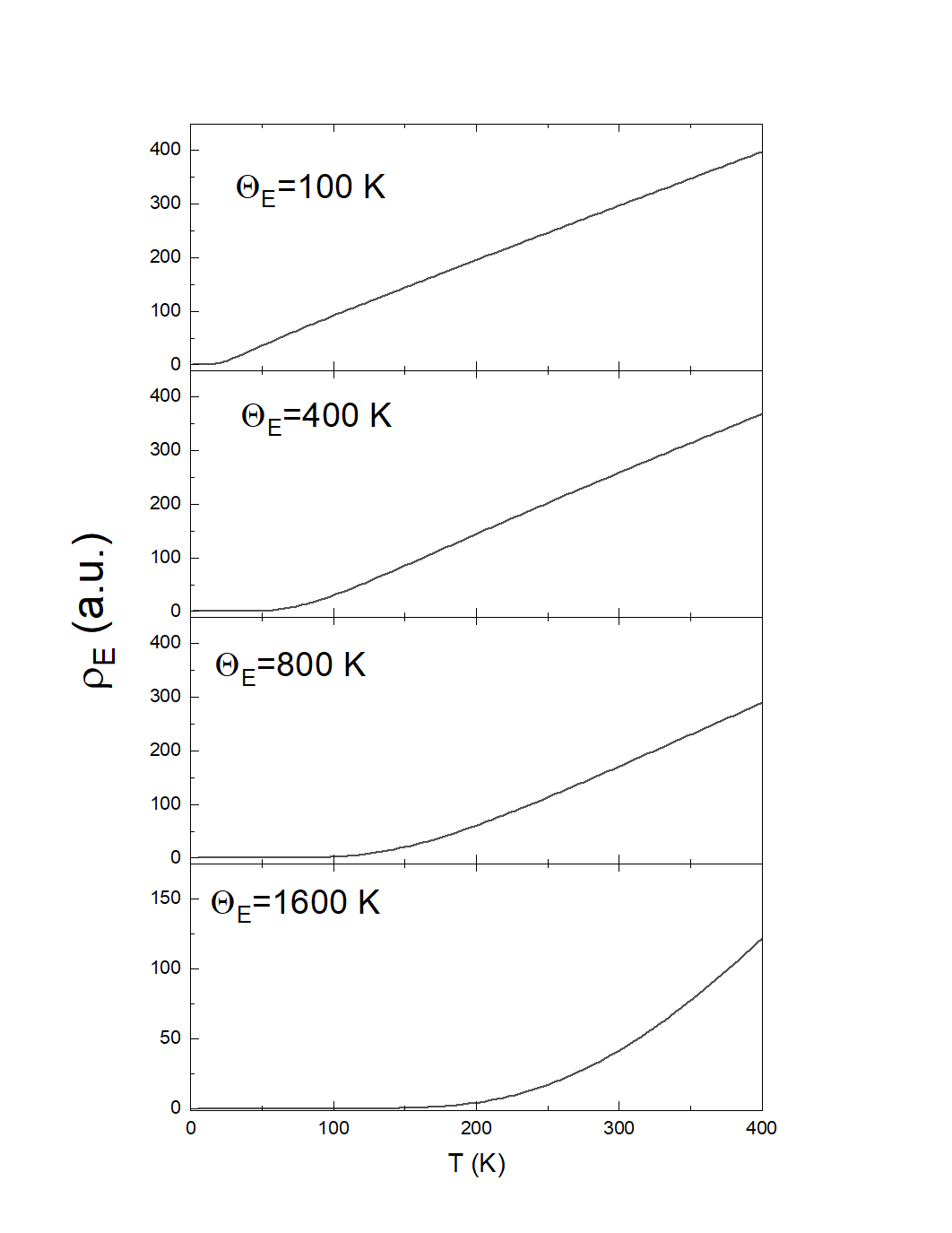}
    \caption{ Temperature dependence of resistivity in Einstein mode model, equation 2, plotted for B=1 and different $\Theta_E$ as indicated in the figure. The evolution suggests that the superlinear dependence is observed for temperatures significantly below $\Theta_E$, and the dependence becomes nearly linear for $T>\Theta_E$, as seen in the top panel. ).    
 }
    \label{fig:Einstein}
\end{figure}

Superlinear temperature dependence of resistivity is usually discussed as an additional contribution to scattering in the Einstein solid model \cite{Einstein1,Einstein2}. This model was invoked to explain the superlinear $\rho(T)$ of MgB$_2$ \cite{MgB2} and LaB$_6$ \cite{Mandrus}. Here, scattering of conduction electrons occurs on optical phonon modes, requiring some finite energy $\Theta_E$ to populate the mode. The dependence is described by the equation

\begin{equation}
\rho_E = B \frac{\Theta_E^2 e^{-\Theta_E/T}}{T~(1-e^{-\Theta_E/T})^2}.
\label{Eq2rhoEnstein}
\end{equation}

\noindent
where the amplitude $B$ is usually used as a fitting parameter. This contribution should be added to $\rho _{BG}$, so that $\rho=\rho_0+\rho_{BG}+ \rho_E$.
The existence of a soft mode in the Raman spectra of BaNi$_2$P$_4$ with a characteristic energy of about 150 cm$^{-1}$ or about 220~K may suggest that scattering on this mode is playing a role in the anomalous temperature dependence of resistivity in BaNi$_2$P$_4$. To gain insight into this possibility, we plot in Fig.~\ref{fig:Einstein} the dependence calculated from the above equation 
for several values of $\Theta_E$. These plots reveal characteristic features of the Einstein mode contribution to resistivity. The resistivity remains temperature independent below $\Theta_E/10$ or so, it is superlinear in the range from $\Theta_E/10$ approximately to $\Theta_E/5$ and it is linear for $T>\Theta_E$. It is clear that scattering invoking modes with $\Theta_E$ as high as 1500 or 1600 K is needed to capture the non-linearity observed in the experiment in BaNi$_2$P$_4$, which is unrealistic.

\begin{figure}
    \centering
    \includegraphics[width = \linewidth]{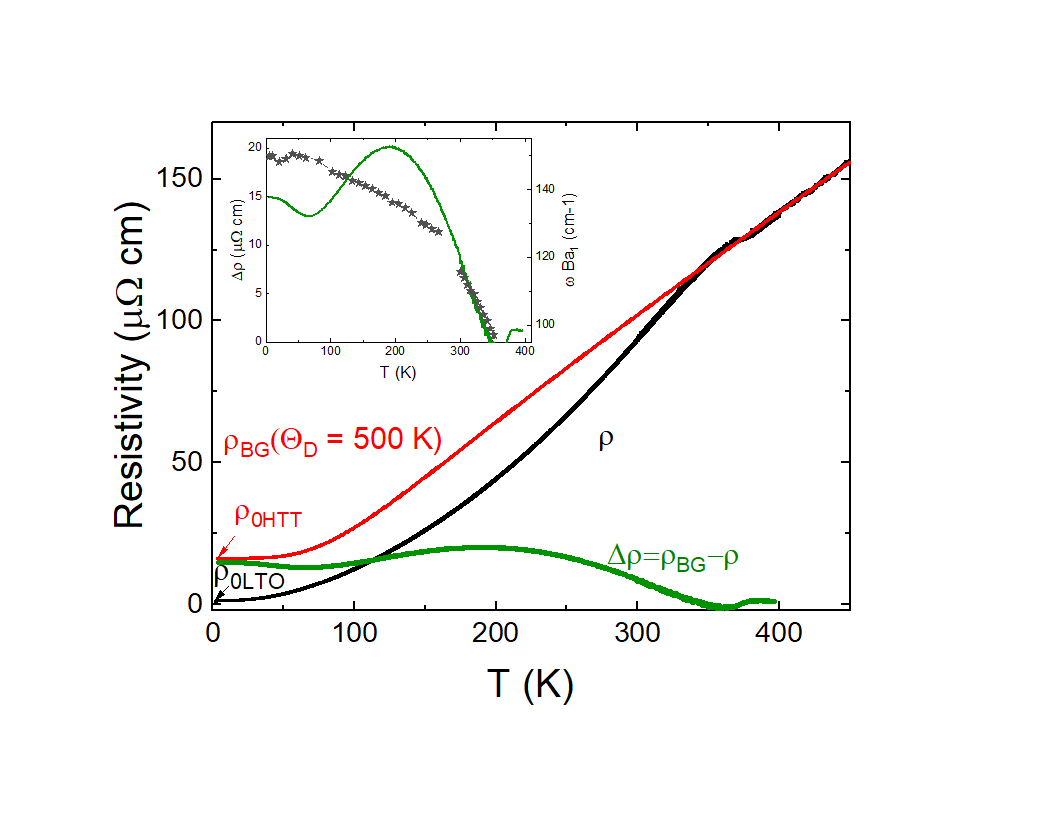}
    \caption{ Temperature dependence of resistivity in BaNi$_2$P$_4$ (black line) and Bloch-Gr\"uneisen fit of the high temperature part with Debye temperature of $\Theta_d=$500~K as suggested from the analysis of high-temperature heat capacity \cite{Novikov} and residual resistivity $\rho_{0HTT}$ significantly higher than $\rho_{0LTO}$ as marked by arrows. Green line is the difference between the actual data and the fit showing contribution equivalent to the loss of residual resistivity between the high temperature HTT phase, $\rho_{0HTT}$, and low temperature LTO phase,  $\rho_{0LTO}$. Inset shows the difference compared to frequency of the Ba$_1$ mode plotted out of scale for comparison (stars).    
 }
    \label{fig:rhoanalysis}
\end{figure}

For the discussion of our results, it is worth bringing to attention other systems revealing strongly non-linear $\rho (T)$ below the metal-metal phase transition.  Notable non-linearity of $\rho(T)$ below the ordering temperature is observed in the two-dimensional superconducting charge density wave materials 2H-TaS$_2$ and 2H-TaSe$_2$ \cite{Naito1982}. The comparison of the temperature dependent resistivity between BaNi$_2$P$_4$ and TaSe$_2$ is made explicitly in Fig.~\ref{fig:Hall} using normalized resistivity and temperature scales. The resistivity curves look quite similar, as shown in the right inset. This is most amazing considering that the Fermi surface of TeSe$_2$ is grossly reconstructed below the transition, as revealed by the sign change and notable increase in the magnitude of the Hall effect in TaSe$_2$, left inset in Fig.~\ref{fig:Hall}.

This result suggests that, in both cases, the dominant effect comes from the variation of scattering. In TaSe$_2$, the $\rho(T)$ at temperatures above the $T_{CDW}$ extrapolates to a very high residual resistivity, suggesting notable charge disorder scattering, which diminishes below $T_{CDW}$ \cite{Naito1982}. In BaNi$_2$P$_4$, the extrapolation of the $T-$linear resistivity from the temperatures above $T_s$ gives a negative intersection on the resistivity axis and a positive intersection on the  temperature axis. This is an expected behavior in BG theory, which is at odds with observations in TaSe$_2$. 

Another material in which a metal-metal transition leads to a non-linear $\rho (T)$ is the organic superconductors $\beta$-(BEDT–TTF)$_2$IBr$_2$ and $\beta$-(BEDT–TTF)$_2$I$_3$ \cite{IBr2}.  These materials display an extended range of $T-$superlinear resistivity behavior, where the resistivity is well-described by the equation $p = p_0 + \alpha T^2$, though the prefactor, $\alpha$, changes at 125~K \cite{IBr2}.   
In organic compounds, this transition is caused by ordering in terminal ethylene groups \cite{Ravy1988}, presumably viewed as disorder by conduction electrons. It is tempting to speculate that the loss of disorder is also responsible for an anomalous resistivity behavior below a metal to metal structural transition in BaNi$_2$P$_4$. The resistivity decrease below $T-$linear extrapolation is not expected in usual metals, but is akin to magnetic metals with strong spin-disorder scattering in the paramagnetic state \cite{Gera,Gera1}, which is lost on magnetic ordering. This is most clearly observed in ferromagnetic materials like LaCrGe$_3$ \cite{LaCrGe3}, in which ordering is not accompanied by the Fermi surface reconstruction. In the right inset of Fig.~\ref{fig:Hall}, we explicitly compare the resistivity of BaNi$_2$P$_4$ and LaCrGe$_3$. The curves indeed bear similarity in their overall dependence. The important relation for understanding this contribution is the direct proportionality between the full magnetic scattering rate (including small angle scattering) and magnetic entropy, as revealed in CeRhIn$_5$ \cite{PaglioneRh}. Although both NMR line splitting and the frequency of the Ba$_1$ mode in Raman scattering indirectly reflect the temperature evolution of the order parameter, they do not directly reflect the entropy change, which is important for scattering.

In Fig.~\ref{fig:rhoanalysis} we explore the idea of loss of scattering below $T_s$. We fit $\rho(T)$ above the transition using a rather high value of $\Theta_D=500$~K, found in heat capacity measurements while taking into account high energy Einstein modes \cite{Novikov}. This fit gives a positive extrapolation of the high temperature part on the resistivity axis (red line in the main panel of Fig.~\ref{fig:rhoanalysis}). The difference between the fit $\rho_{BG}$ and experimental $\rho(T) $ is shown as a green line in the main panel of Fig.~\ref{fig:rhoanalysis}. This difference resembles the behavior of the structural order parameter, inset in Fig.~\ref{fig:rhoanalysis}. 
We thus conclude that the Ba atom rattling mode in the cage in the high temperature tetragonal phase adds an additional local contribution with $\rho_{0HTT}>\rho_{0LTO}$ and adds a constant offset to the high temperature part of the resistivity.


\section{Conclusions}

In summary, we found that the anomalous superlinear temperature-dependent resistivity of BaNi$_2$P$_4$ is observed only below the tetragonal-to-orthorhombic transition at around 376~K. The transition does not notably affect carrier density, as evidenced by the validity of the Matthiessen rule and the temperature-independent Hall effect. The behavior above the transition follows the expectations of Bloch-Gr\"uneisen theory for a normal metal with $T-$linear resistivity, extrapolating to a positive offset on the resistivity axis. We attribute the down-ward deviation from linear dependence below $T_s$ to a loss of additional residual resistivity contribution above $T_S$.  We conclude that the  super-linear resistivity variation is caused by the anomalous temperature-dependence of the scattering rate, most likely related to the build-up of the structural order parameter.

\begin{acknowledgments}
The authors thank P.~Alemany for insightful discussion. 
This work was supported by the U.S. Department of Energy (DOE), Office of Science, Basic Energy Sciences, Materials Science and Engineering Division. Ames Laboratory is operated for the U.S. DOE by Iowa State University under contract DE-AC02-07CH11358. The authors acknowledge support from the EMIR\&A French Network (FR CNRS 3618) on the platform SIRIUS. Some of the computation used resources of the National Energy Research Scientific Computing Center (NERSC), a DOE Office of Science User Facility. K.K. and G.V. acknowledge support by the U.S. Department of Energy, Office of Basic Energy Sciences, Division of Materials Science and Engineering, grant DE-SC0022288. R. G. thanks Alexandr Alekhin and Yann Gallais for giving access to the AFM-Raman platform of Universit\'e Paris Cit\'e which is supported by an IdEx Grant, ANR-18-IDEX-0001. OD acknowledges support 
from the U.S. Department of Energy, Office of Science, Basic Energy Sciences, Materials Sciences and Engineering
Division, under Award No. DE-SC0019978.

E.H.K., M.A.T. and E.I.T. performed electrical resistivity and Hall effect measurements as a function of electron irradiation.
E.H.K. and E.I.T. conducted polarized light imaging of domains.
S.L.B. and P.C.C. performed high temperature magnetization measurements. 
Q.P.D. and Y.F. performed NMR measurements.
L.L.W. performed DFT band structure calculations and Hall effect calculations.
M.K. and R.G. performed electron irradiation.
R.G. performed Raman scattering measurements.
J.N. and O.D. conducted inelastic neutron scattering measurements.
G. V. and J.W. grew single crystals of BaNi$_2$P$_4$ and BaCu$_2$P$_4$ and performed
their structural characterizations,
K.K., R.P. and M.A.T. designed the project.
M.A.T., E.H.K., R.G. Y.F., K.K. and R.P. drafted the manuscript with inputs from all authors.

\end{acknowledgments}



\end{document}